\documentclass[sigconf]{acmart}
\AtBeginDocument{%
  }

\setcopyright{acmlicensed}
\copyrightyear{2026}
\acmYear{2026}
\acmDOI{10.1145/3785022.3785075}
\acmConference[LAK'26]{The 15th International Learning Analytics and Knowledge Conference (LAK 2026)}{April 27-May 01,
  2026}{Bergen, Norway}

\usepackage{minted}
\usepackage{caption}
\usepackage{graphicx}

\newcommand{\fig}[1]{Fig.~\ref{fig:#1}}

\newcommand{\myfigspaceend}{}
\newcommand{\startmycapspace}{}
\newcommand{\icaption}[1]{\startmycapspace\caption{\sl #1}}
\newcommand{\dcaption}[1]{\icaption{#1}\Description[#1]{#1}}

\begin{document}

\title{Changes in Coding Behavior and Performance Since the Introduction of LLMs}

\author{Yufan Zhang}
\orcid{0009-0004-3791-7311}
\affiliation{%
  \institution{Carnegie Mellon University}
  \city{Pittsburgh}
  \country{U.S.A.}
}
\email{yufanz@cmu.edu}

\author{Jaromir Savelka}
\orcid{0000-0002-3674-5456}
\affiliation{%
  \institution{Carnegie Mellon University}
  \city{Pittsburgh}
  \country{U.S.A.}
}
\email{jsavelka@cmu.edu}

\author{Seth Copen Goldstein}
\orcid{0000-0003-1512-0446}
\affiliation{%
  \institution{Carnegie Mellon University}
  \city{Pittsburgh}
  \country{U.S.A.}
}
\email{seth@cmu.edu}

\author{Michael Conway}
\orcid{0009-0007-1974-1596}
\affiliation{%
  \institution{Udacity}
  \city{Chapel Hill}
  \country{U.S.A.}
}
\email{michael.conway@udacity.com}

\renewcommand{\shortauthors}{Zhang et al.}

\begin{abstract}

The widespread availability of large language models (LLMs) has changed how students engage with coding and problem-solving. While these tools may increase student productivity, they also make it more difficult for instructors to assess students' learning and effort. In this quasi-longitudinal study, we analyze five years of student source code submissions in a graduate-level cloud computing course, focusing on an assignment that remained unchanged and examining students' behavior during the period spanning five semesters before the release of ChatGPT and five semesters after.

Student coding behavior has changed significantly since Fall 2022. The length of their final submissions increased. Between consecutive submissions, average edit distances increased while average score improvement decreased, suggesting that both student productivity and learning have decreased after ChatGPT's release. Additionally, there are statistically significant correlations between these behavioral changes and their overall performance. Although we cannot definitively attribute them to LLM misuse, they are consistent with our hypothesis that some students are over-reliant on LLMs, which is negatively affecting their learning outcomes.  Our findings raise an alarm around the first generation of graduates in the age of LLMs, calling upon both educators and employers to reflect on their evaluation methods for genuine expertise and productivity.

\end{abstract}





\maketitle

\section{Introduction}

Rapidly increasing capabilities of large language models (LLM)
continue challenging existing practices in professional software
development settings as well as in computer science education. The now
well-established mode of interaction where an LLM is manually
triggered to complete, create, or re-factor individual pieces of code
is recently being replaced by workflows involving tools such as Claude
Code\footnote{\url{https://claude.com/product/claude-code}} or Gemini
CLI\footnote{\url{https://cloud.google.com/gemini/docs/codeassist/gemini-cli}}
that are often referred to as agents. Although agentic frameworks
handle increasingly more complex tasks such as planning, testing, and
refactoring, human developers still play a crucial role in defining
goals, guiding the process, and reviewing and fixing outcomes. At the
extreme end, vibe coding is being used, where high-level natural-language directives
orchestrate end-to-end software creation~\cite{ray2025review}. It appears that over a short period of
several years, software development has transitioned from a state where
most of the code is written by humans to one where the great
majority of code is written by AI agents. At the same time, human
supervision and intervention remain and will likely continue to remain
indispensable. However, as there are increasingly fewer natural
incentives for programmers to directly write and read code, a
critical question emerges: How can we ensure that students and future
software engineers develop and systematically hone their programming
expertise?

To even attempt to answer the question posed above we need to explore,
quantitatively, how students are using LLMs in their work and how the
use of an LLM is changing outcomes.  Indeed, there has been a growing
body of scholarship focused on how software developers and students
use and perceive LLMs to assist them in carrying out their
work. Although it is indisputable that students and software
developers increasingly integrate LLM assistants into their workflows,
empirical evidence on how this is happening, to what extent, and what
effects it has remains scattered and limited. There are
conflicting accounts ranging from how LLMs improve student
learning~\cite{liffiton2023codehelp,sheese2024patterns} to how LLMs
harm students who over-rely on them~\cite{prather2024widening}. There may also be a mismatch between students' perceptions of increases in their productivity due to LLM usage and actual changes in their productivity. It has been shown that while
experienced software developers may consider the use of LLM-powered
assistants as boosting their productivity, in fact the effect may be
quite opposite~\cite{becker2025measuring}. Hence, there is an urgent
need for studies showing how the coding behaviors of students and
software developers have been evolving in this rapidly transforming
landscape and what effects these changes bring about.

This paper analyzes students' source code submissions in a graduate-level computer
science course in cloud computing at Carnegie Mellon University in the United States, over 10
semesters between Fall 2020 and Spring 2025, i.e., precisely the
period from several years before and several years after the
introduction and mass popularization of LLM-powered coding assistants
and dialog systems. The purpose of this paper is to understand how
LLMs are impacting our students.  We use code submissions and how they
are made as a proxy to students' learning behavior over the studied
period.  To that end, we employ and analyze a data set comprising 2,066
student submissions from the 10 semesters. The immediate insight that
this study offers is that since the introduction of LLMs students
approach to solving the task has changed significantly.  They make
more changes between consecutive submissions, they submit slightly more often, and they
produce significantly more bloated solutions, as shown in Fig.~1.  We show that these behavioral changes, though sometimes sufficient for completing programming tasks, may be translating into reduced fulfillment of the learning objectives while masking it under improved performance.

\begin{figure}[!t]
    \centering
    \includegraphics[width=\columnwidth]{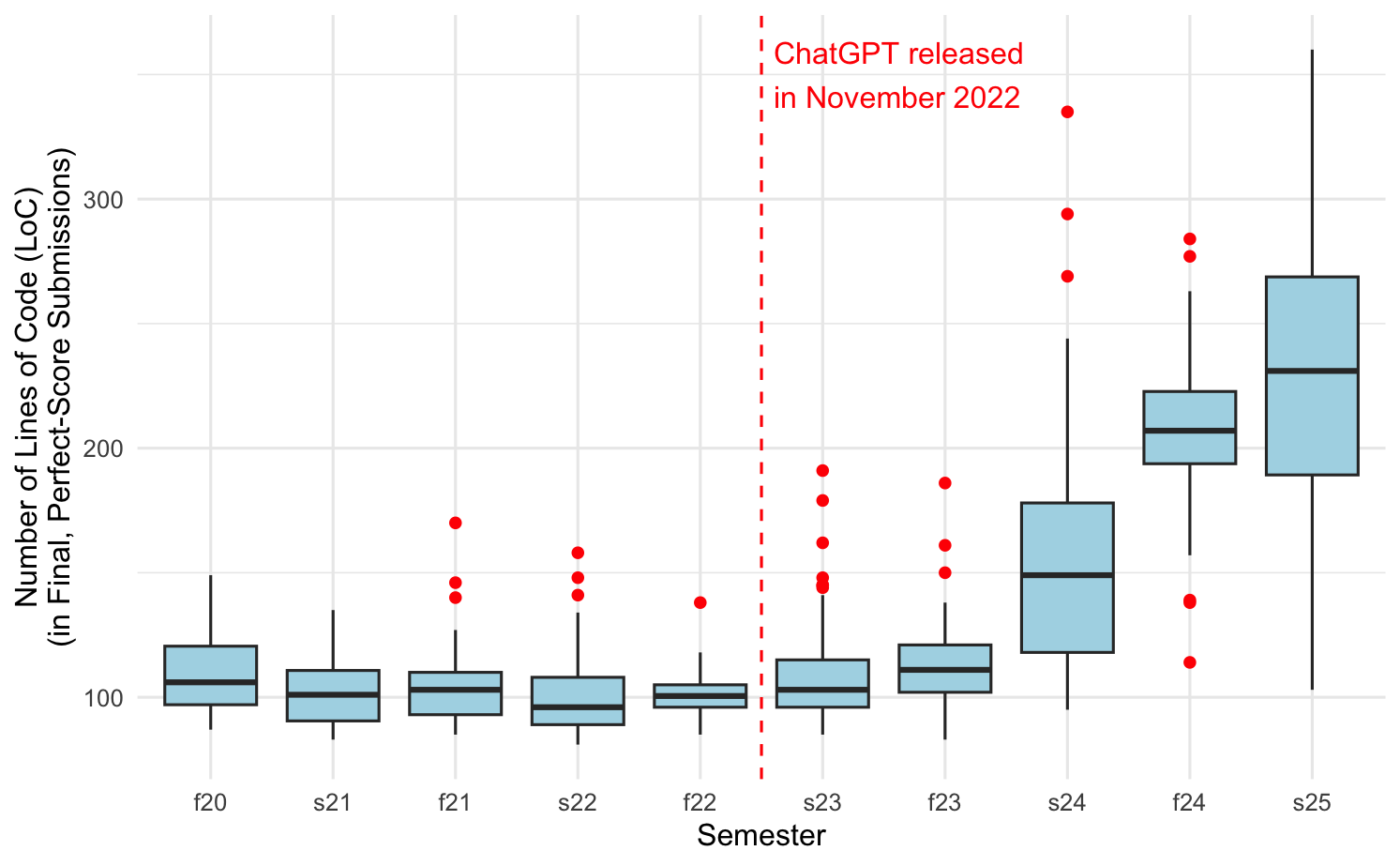}
    \dcaption{Lines of code in final, perfect-score submissions increased significantly since Fall 2022. Note: the y-axis is not zero-indexed.}
    \label{fig:final_submission_length}
\myfigspaceend
\end{figure}

We address the following research questions in our work:

\begin{enumerate}
\item[RQ1] How has coding behavior in individual assignments changed
  among students since LLM coding assistants and dialog systems became
  widely available?
\item[RQ2] Are changes in individual coding behavior associated with
  changes in learning efficiency and learning outcomes?
\end{enumerate}

To our best knowledge, this is the first comprehensive study that provides empirical evidence on changes in coding behavior, learning outcomes, and learning paths over 10 consecutive semesters of observations.

We hope the quantitative analysis of student behavior presented in this
paper will spur a deeper conversation about how our educational
approach should shift in the age of AI.  A useful lens for thinking
about the needed transformation comes from the notion of centaurs in
chess.  The centaur, a team where humans and machines work together,
outperforms either working on its own~\cite{dk21,sl25}. The lesson from
that domain is not simply that tools can make people faster, but that
they fundamentally change what humans need to do, how they do it, and
how we measure excellence. Software development is undergoing a
similar shift. As AI agents take on larger portions of planning,
scaffolding, testing, and implementation, the human role may be less
about tactics (i.e., producing code) and more about strategy (i.e.,
defining abstractions, structuring complex systems, integrating
components, and critically evaluating automated output). This has
profound implications for project-based courses in computing
education: our curricula must help students learn how to collaborate
with intelligent tools, not just use them; our assessments must evolve
to measure what is important for the human part of the centaur.

A key insight that propelled this work is an observation of a common behavior of LLMs. Specifically, when they are prompted for editing text --- whether it is in a natural language or a programming language --- they have a tendency to create and update parts of the text they have not been explicitly prompted to. Examples include formatting tokens, emojis, and additional documentations. A side effect of this behavior is that the source code gets longer faster, and the edit distance increases significantly. This discovery serves as the basis of our speculation of the extent to which a student may have utilized LLMs for code generation.

\section{Related Work}

\subsection{Coding in the Age of LLMs}
The public release of ChatGPT~\cite{chatgpt} and other LLMs, especially those accessible through chat interfaces or integrated into development environments, has changed how students and professionals write code. LLMs enable users to generate, revise, and explain code on demand, often bypassing the need to deeply understand the underlying logic or syntax. With LLMs, students might copy and paste code snippets into their editors, modifying them as needed. In a study where undergraduates annotated their code to indicate which parts were generated by LLMs, 40.5\% of teams reported LLM usage for their semester-long software engineering project~\cite{Rasnayaka24}. Other studies find that students use LLMs not only to generate code, but also to interpret error messages and auto-grader feedback~\cite{sun2024would}. In fact, ``reading feedback'' and ``understanding Python code'' were among the most common use cases after code generation itself~\cite{sun2024would}. In a controlled experiment with screen monitoring, students chose ChatGPT over traditional online resources such as Google and Stack Overflow~\cite{xue24}. LLMs are also used in earlier stages of professionals' software development life-cycle, including planning, design, and testing~\cite{Khojah24}. However, reliance on LLMs introduces new risks: students do not always communicate specifications effectively, and generated code may fail to meet their intentions~\cite{Nguyen24,sun2024would}. As project complexity increases, students often reduce LLM usage~\cite{Rasnayaka24}, and some prefer human assistance when available~\cite{lyu2024evaluating}.


\subsection{Grading for CS Education in the Age of LLMs}
The widespread availability of LLMs complicates the assessment of student learning. While students may complete tasks more quickly or more accurately with LLM assistance, this does not always reflect internalized understanding. Studies on the relationship between LLM usage and academic performance show mixed results~\cite{wang2025effect,ramirez2025understanding}. Some studies report learning gains when LLMs are used in tutoring roles~\cite{lyu2024evaluating}, while others find no significant performance difference between users and non-users~\cite{sun2024would,kosar2024computer}.
Behaviorally, students who avoid ChatGPT tend to debug code independently~\cite{sun2024would}. Meta-analyses suggest LLMs may reduce task time and improve assignment scores, but often fail to improve perceived learning~\cite{alanazi2025influence}. One study of students building React apps found a negative correlation between LLM reliance and final grades~\cite{jovst2024impact}, highlighting the risk of over-reliance. Moreover, attaining high scores does not necessary entail meeting learning objectives if the artifacts themselves are co-authored by LLMs. We contribute empirical evidence that also shows a negative correlation between potential LLM usage and learning outcomes, and we do so via a longitudinal lens.

\section{Data}

We analyzed data from 10 consecutive semesters of the Cloud Computing course at Carnegie Mellon University, taken by both undergraduate and
graduate students between Fall 2020 and Spring 2025. With ChatGPT
being introduced in November 2022, our observations include 5
pre-ChatGPT semesters and 5 post-ChatGPT semesters. We sometimes denote semesters by concatenating the season and the
year. For example, f20 denotes Fall 2020 and s25 denotes Spring 2025.

The course is challenging, requiring upwards of 12--15 hours of
active effort per week, with some students reportedly spending 20--40
hours in some weeks. Students taking the course are assumed to be
proficient in Java and/or Python, and usually have already taken an
introductory computer systems course in C. Intensive programming is
part of the course but not among the learning objectives which
encompass cloud computing basics (elasticity and containerization),
storage, and applications (iterative processing, stream processing,
and machine learning).

The course has many graded components, including a sequence of 6
individual projects (each consisting of a sequence of 3--7 tasks), a
team project (which consists of 3 phases), and 11 weekly
quizzes. A semester usually lasts 14 weeks excluding holidays and exam weeks. Collaboration and AI-assisted coding are prohibited for individual projects and quizzes. In this study, we focus on one specific task from one
specific individual project and the team project. For both of them, we consider only the auto-graded components, where students receive near constant feedback on the source code they have submitted based.

We selected the task within the individual project (which we denote as the PageRank task) because it remained essentially the same between Fall 2020 and Spring 2025, with no changes to the specification and minimal changes to the handout, allowing us to easily compare the starter code (which contains 60 lines written in Bash and Scala) and student solutions from all semesters. As fixed set of tests are run against each submission such that the same source code will get the same score when submitted at different times, and if a submission gets a different score it must be meaningfully different from another submission. Other tasks in the course are not selected for this study primarily because they have undergone non-trivial changes over the course of 10 semesters or have a much smaller solution space (such as Dockerfiles, Kubernetes YAML files, and MySQL queries) that inherently limit variances among different students' solutions. For the PageRank task, students analyze a Twitter social graph and implement the PageRank algorithm (in as few as 40 new lines of code, as shown in \fig{final_submission_length}) to find the most influential users on Twitter. Students can submit their source code to be evaluated by auto-graders as often as they like before the deadline, immediately receiving a score and feedback, which they could use to improve their solution. Only the score of their final submission is recorded, though all submissions are logged.

We also included students' performance in the team project (which we denote as TP) as an alternative measure of one's learning outcome and productivity, albeit confounded by their teammates' contributions. Students work in teams of 2 or 3 over 10 weeks to develop and deploy a full-fledged web service. They do so from scratch, provided only the specification of the web service without any starter code or template. Instead of submitting their source code, they deploy their web service on Amazon Web Service (AWS) and submit an endpoint to their service, which is stress tested by load generators and graded based on correctness and performance. Unlike for the PageRank task, students receive auto-graded scores and feedback for the team project only at the end of each of the three phases (6 weeks in, 8 weeks in, and 10 weeks in) when load generators are run, mimicking production workflows. Students' source code for the team project is not considered in this study, as there are no discrete timestamps (like there are for the PageRank task) between which edit distances can be reliably measured.

\begin{figure}[tb]
  \centering
  \includegraphics[width=\linewidth]{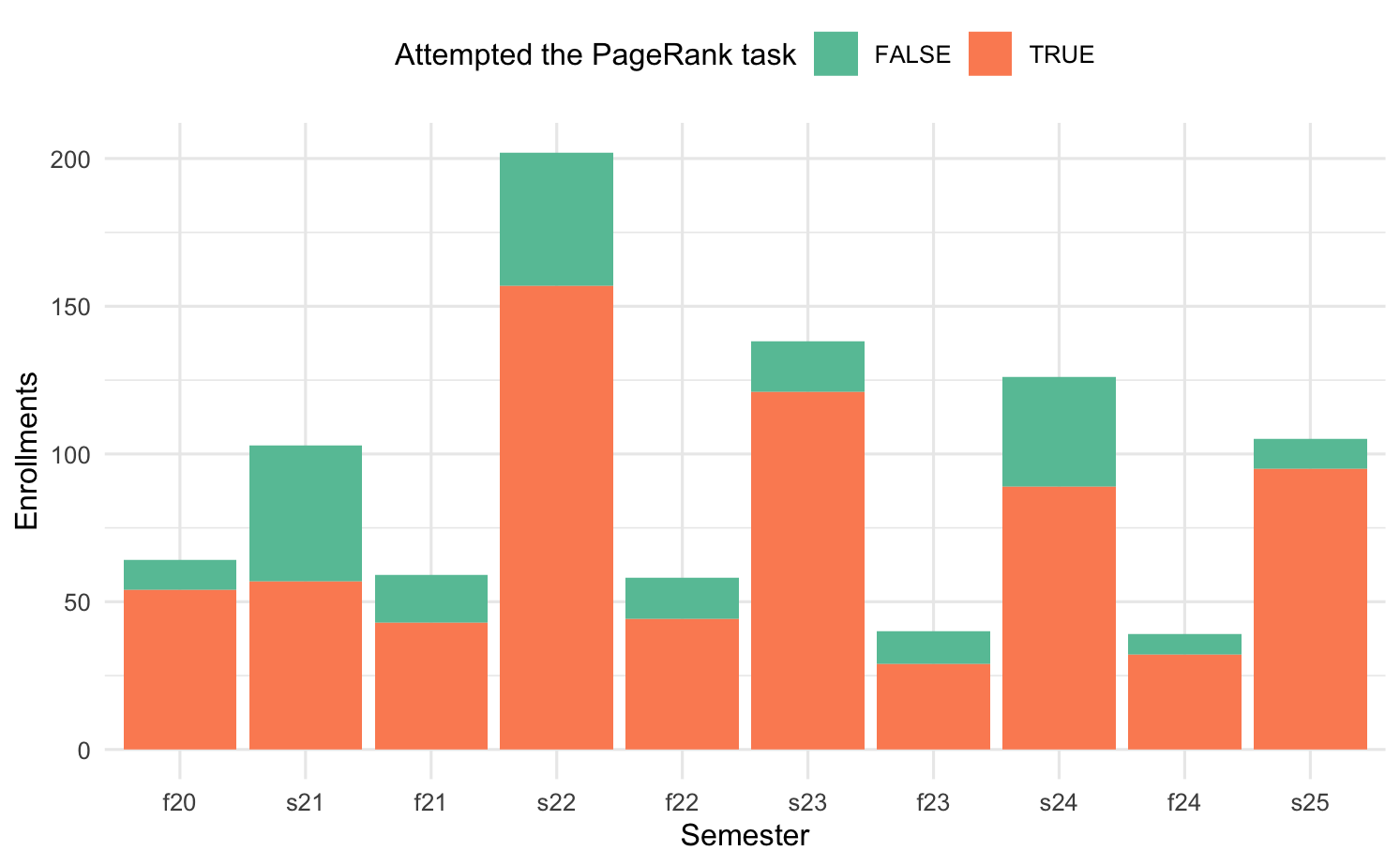}
  \dcaption{Enrollments by semester, split by participation in the PageRank task.}
  \label{fig:enrollments_by_participation_in_ipt}
\end{figure}

\begin{figure}[tb]
  \centering
  \includegraphics[width=\linewidth]{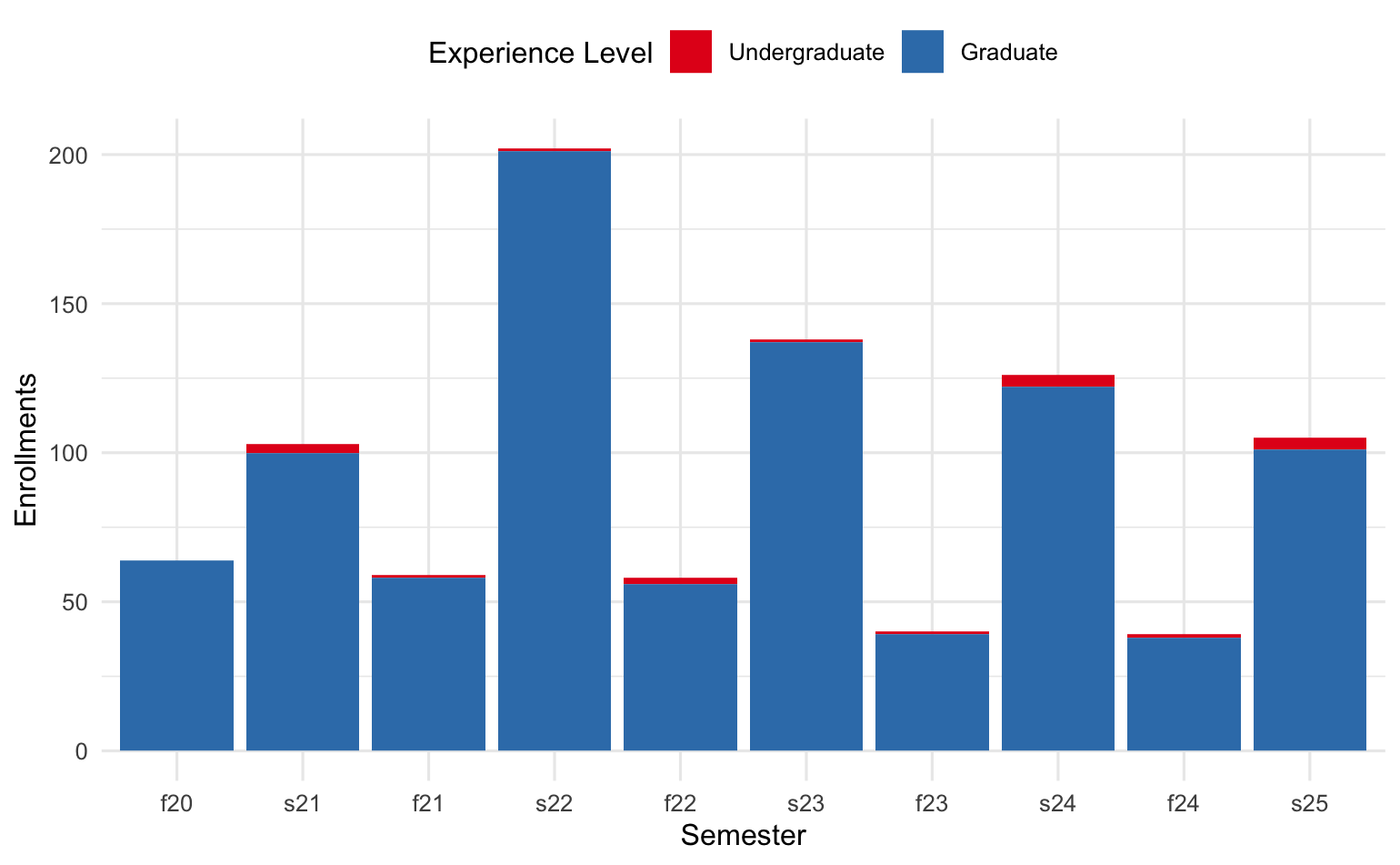}
  \dcaption{Enrollments by semester, split by experience Level.}
  \label{fig:enrollments_by_experience_level}
\end{figure}

There are 936 enrollments in total between Fall 2020 and Spring 2025,
but only 721 enrollments (718 unique students) attempted the PageRank task, as
shown in \fig{enrollments_by_participation_in_ipt}. This is because out of the 6 individual projects in the course, only the 5 highest-scoring projects are used in the calculation of each student's final grade, and some students may choose to drop the project in which the PageRank task exists. If the same student enrolls in the course in different semesters, they are
considered separate enrollments. Almost all
enrollments are graduate students, as shown in
\fig{enrollments_by_experience_level}. For this study, we only consider
the 721 enrollments who attempted the PageRank task, which means they made at least one submission for the task. They are represented by the orange-colored bar segments in \fig{enrollments_by_participation_in_ipt}.

\section{Methods}

\subsection{Coding Behavior}

To measure changes in coding behavior, we downloaded the source code from all student submissions for the PageRank task between Fall 2020 and Spring 2025 (concatenating the Bash and Scala files such that there is a single text file representing each submission made by each student), and defined the following metrics:

\begin{itemize}
    \item \texttt{Number of Submissions (Attempts)}: The number of times a student submits their source code for the PageRank task.
    \item \texttt{Total Edit Distance}: We measure edit distances
      based on the number of lines changed between submissions, using
      Myers algorithm which is the default algorithm used by {\tt git diff}~\cite{myers}. The edit distance for the initial
      submission by a student is computed with respect to the starter code that is provided to all students. The
      \texttt{Total Edit Distance} is the sum of the edit distances over the
      sequence of all submissions a student makes for the task.
    \item \texttt{Average Edit Distance}: \texttt{Total Edit Distance} divided by the \texttt{Number of Submissions}.
\end{itemize}

\subsection{Performance}

To measure changes in performance for each student, we define the following metrics.

\begin{itemize}
    \item \texttt{Task Score}: The auto-graded score a student receives for their final submission for the PageRank task. The \texttt{Task Score} shows the direct effect of each student's approach to the task.
    \item \texttt{IP Score}: The sum of auto-graded scores a student receives for their top 5 out of 6 individual projects in the course, which may be a different mix for each student. \texttt{IP Score} shows how well a student has likely met the learning objectives of the course which encompass multiple topic areas and coding projects.
    \item \texttt{TP Score}: The sum of the auto-graded scores a student receives in all three phases of the team project. The \texttt{TP Score} shows how well a student has likely applied their learning to building a production web service in a team, even though contributions from the team members may be unequal.
\end{itemize}



\section{Results}\label{sect-results}

\begin{figure}[tb]
        \centering
        \includegraphics[width=\linewidth]{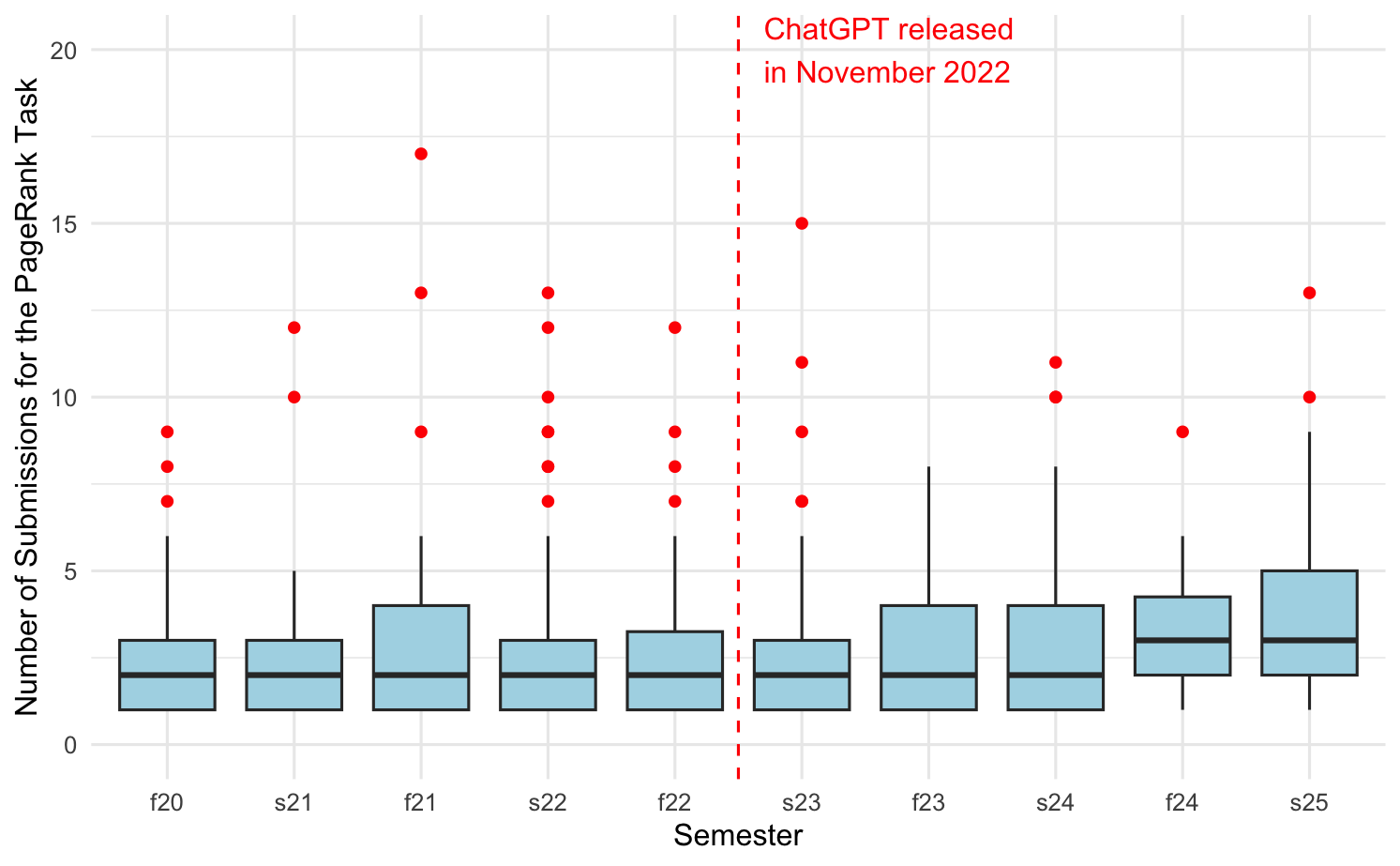}
        \dcaption{Number of submissions per student ticked up slightly since s23.}
        \label{fig:num_of_ipt_submissions}
\end{figure}

We examine changes in coding behavior (visualized in \fig{num_of_ipt_submissions}, \fig{tot_edit_distance}, \fig{avg_edit_distance}) and performance (\fig{ipt_scores}, \fig{ip_scores}, \fig{tp_scores}), as well as associations between them (\fig{avg_edit_effectiveness}, \fig{prop_submissions_score_deltas}, \fig{ip_score_vs_ipt_dist}, \fig{tp_v_avg_edit_dist}).

\subsection{Changes in Coding Behavior}

There are 2,066 submissions in total from the 721 enrollments. Most students attempted the task fewer than 5 times, where the median is 2
submissions; however, there is a slight but noticeable increase in \textbf{\texttt{Number of Submissions}} starting in f23, and in its median starting in f24, as shown in \fig{num_of_ipt_submissions}. For 8 consecutive semesters leading up to s24, half the students made no more than 2 attempts, but by f24 it had increased to no more than 3 attempts, suggesting that students required more attempts to complete the task after the introduction of LLMs .

\begin{figure}[tb]
        \centering
        \includegraphics[width=\linewidth]{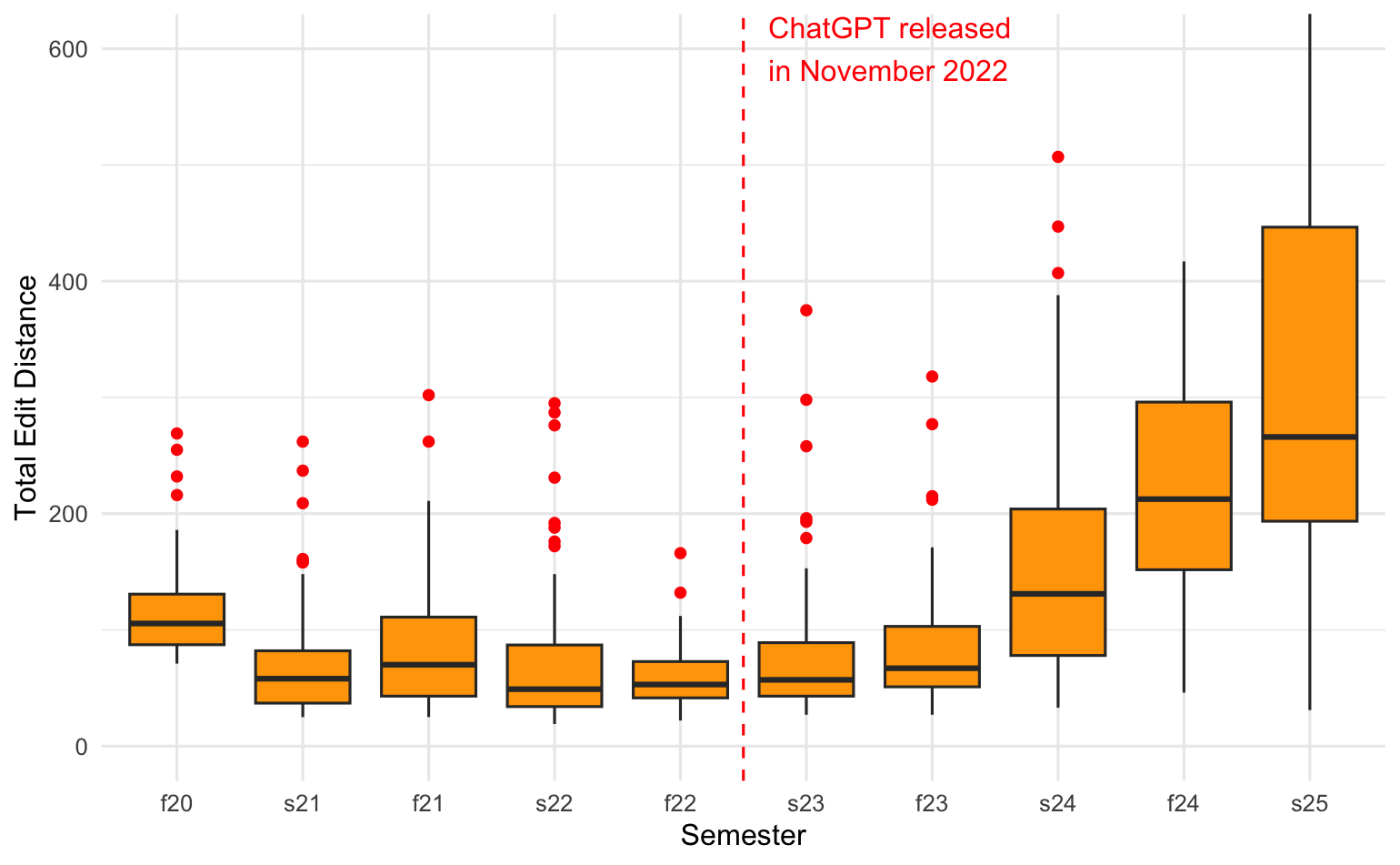}
        \dcaption{Total edit distance increased since s23}
        \label{fig:tot_edit_distance}
\end{figure}

\begin{figure}[tb]
        \centering
        \includegraphics[width=\linewidth]{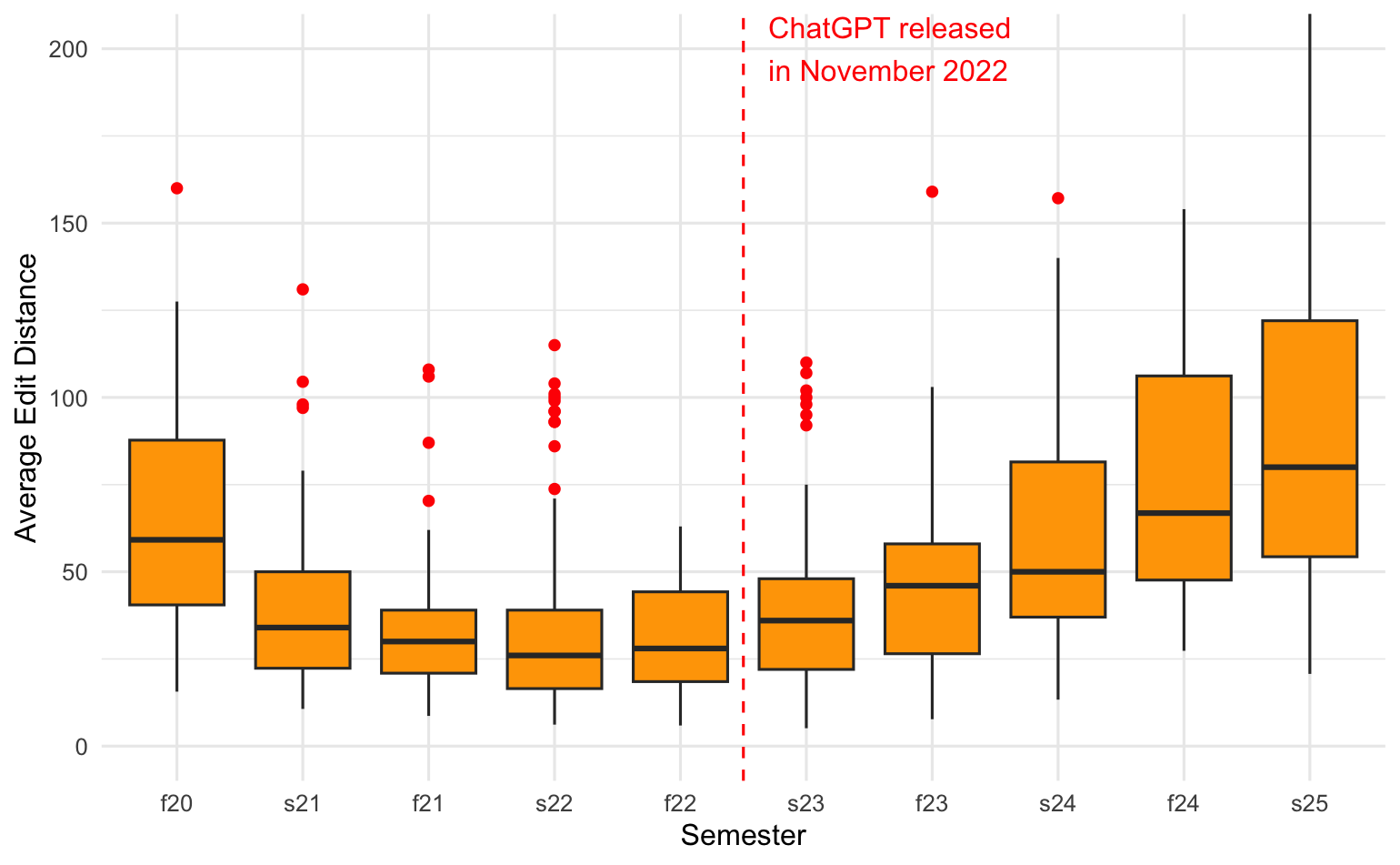}
\dcaption{Average edit distance increased since s23}
        \label{fig:avg_edit_distance}
\end{figure}

\textbf{\texttt{Total Edit Distance}} and \textbf{\texttt{Average Edit Distance}} increased significantly starting in s23, as shown in \fig{tot_edit_distance} and \fig{avg_edit_distance}, where the former increased by almost a factor of 10 between Fall 2022 and Spring 2025, and the latter almost tripled over the same period. The observations are consistent with our hypothesis that students are prompting LLMs to generate the code for them, and copying-and-pasting the code into their solutions. Our hypothesis is based on observations that LLMs tend to change existing text and code in ways they are not requested to change. Whether or not they lead to performance gains (which, as we show in the next subsection, they do not), higher edit distances (unless for better styles and documentations) require longer code review and can negatively affect productivity.

\subsection{Changes in Performance}

\begin{figure}[tb]
        \centering
        \includegraphics[width=\linewidth]{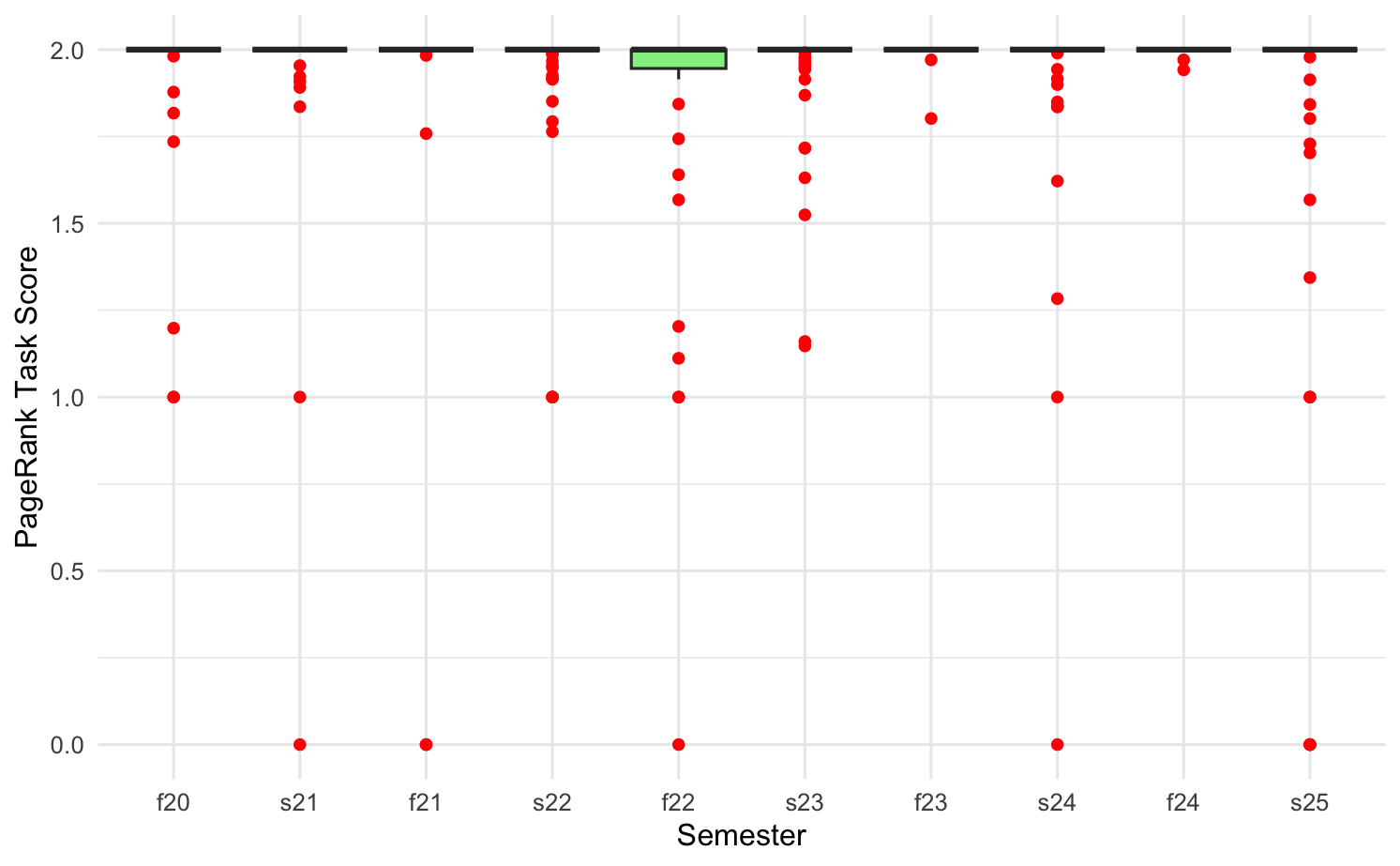}
        \dcaption{Almost all students received a full score for PageRank}
        \label{fig:ipt_scores}
\end{figure}

In all semesters except f22, nearly all students who attempted the PageRank task ultimately received a perfect \texttt{Task Score}, as shown in \fig{ipt_scores}. This suggests that task performance alone cannot adequately reflect the efficacy of different approaches to the task, including potential LLM usage. There is little variance in task performance that could plausibly be explained by differences in coding behavior. These results may also indicate that the task is too easy.

\begin{figure}[tb]
  \centering
  \includegraphics[width=\linewidth]{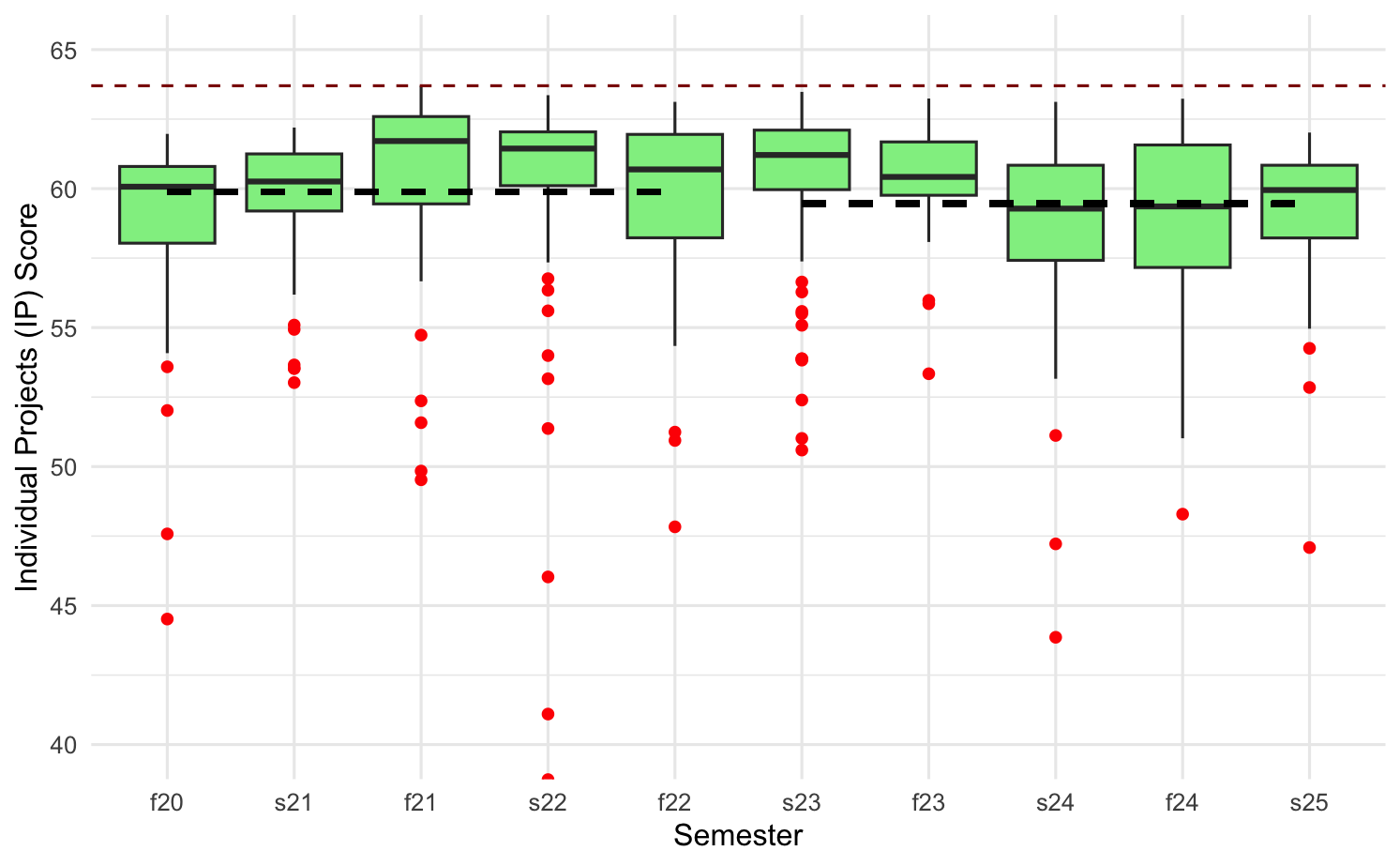}
  \dcaption{Individual Projects (IP) scores by semester.}
  \label{fig:ip_scores}
\end{figure}

There are no discernible trends in \textbf{\texttt{IP Score}}, except for a slight decrease in the average \texttt{IP Score} before and after f22, as shown in \fig{ip_scores}. As discussed in the next subsection, while the decrease could be attributed to increasing over-reliance on LLMs, our observations could suffer from a ceiling effect, where most students have already been acing the course. In fact, the median \texttt{IP Score} continues to hover around 60 out of a maximum possible score of 63.7 (or 94\%) suggesting that half the students continue to ace the individual projects overall.

\begin{figure}[tb]
  \centering
  \includegraphics[width=\linewidth]{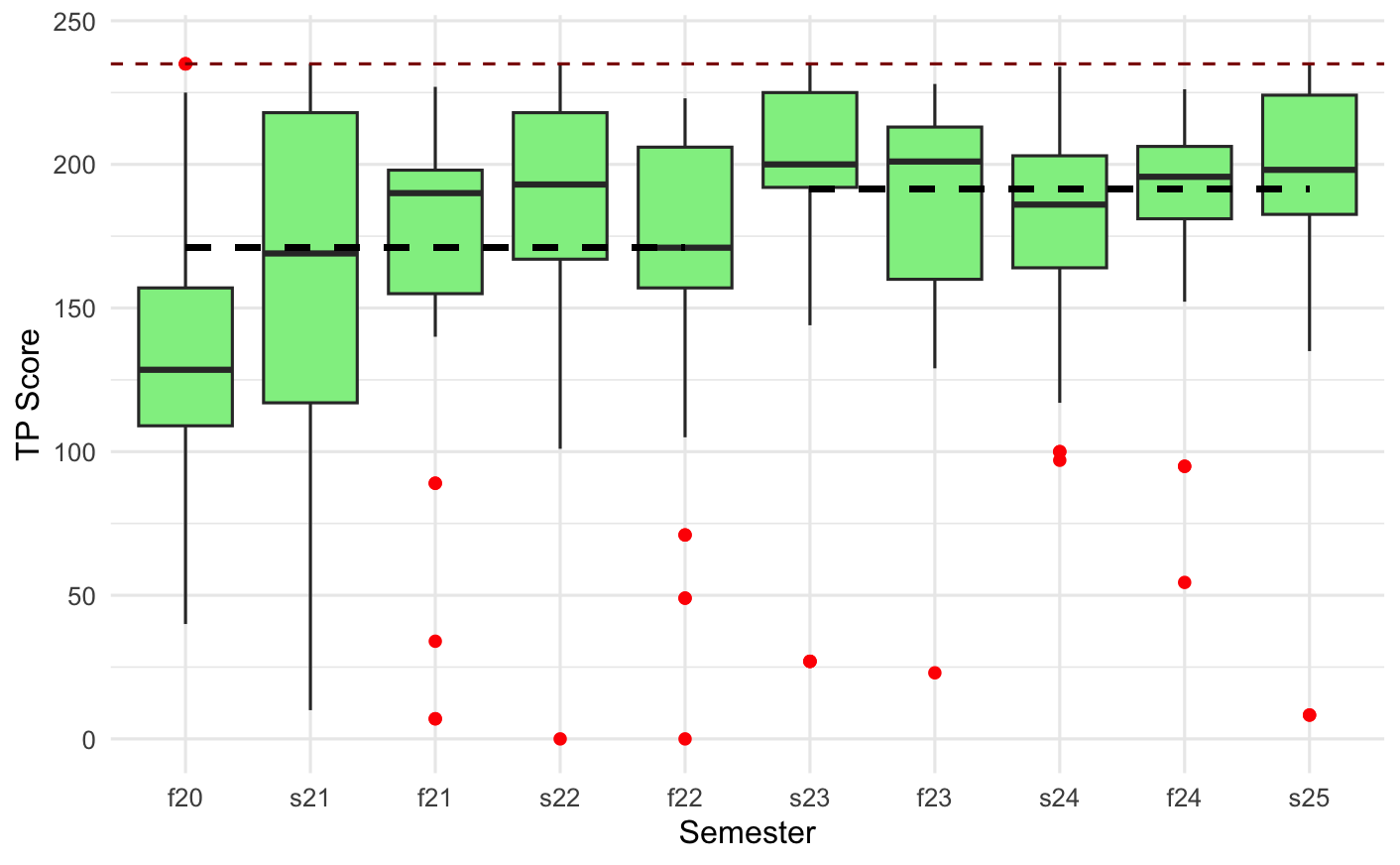}
  \dcaption{Team Project (TP) scores by semester. The low scores in Fall 2020 (f20) are likely an outlier, due to COVID}
  \label{fig:tp_scores}
\end{figure}

The average \textbf{\texttt{TP Score}} noticeably increased after f22, as shown in \fig{tp_scores}. Where the median used to fluctuate between 150 and 200 (out of a maximum possible score of 235), it seems to have been hovering around 200 (or 85\%) since s23. Though it is unclear if the trend actually exists or what may have caused the trend, our hypothesis is that AI assisted coding has made it easier for individuals who are competent and committed to carry their team, utilizing AI chat bots as if collaborating with an additional team member that is not officially recognized as a contributor, assisting not only with code generation but also thought processes. We plan to test that hypothesis in a future study. The \texttt{TP Score} in f20 is remarkably low; we think it is because it was the first semester running entirely in COVID lockdown, where students were not used to collaborating without face-to-face interactions.

\subsection{Correlating Behavior With Performance}

\begin{figure}[tb]
        \centering
        \includegraphics[width=\linewidth]{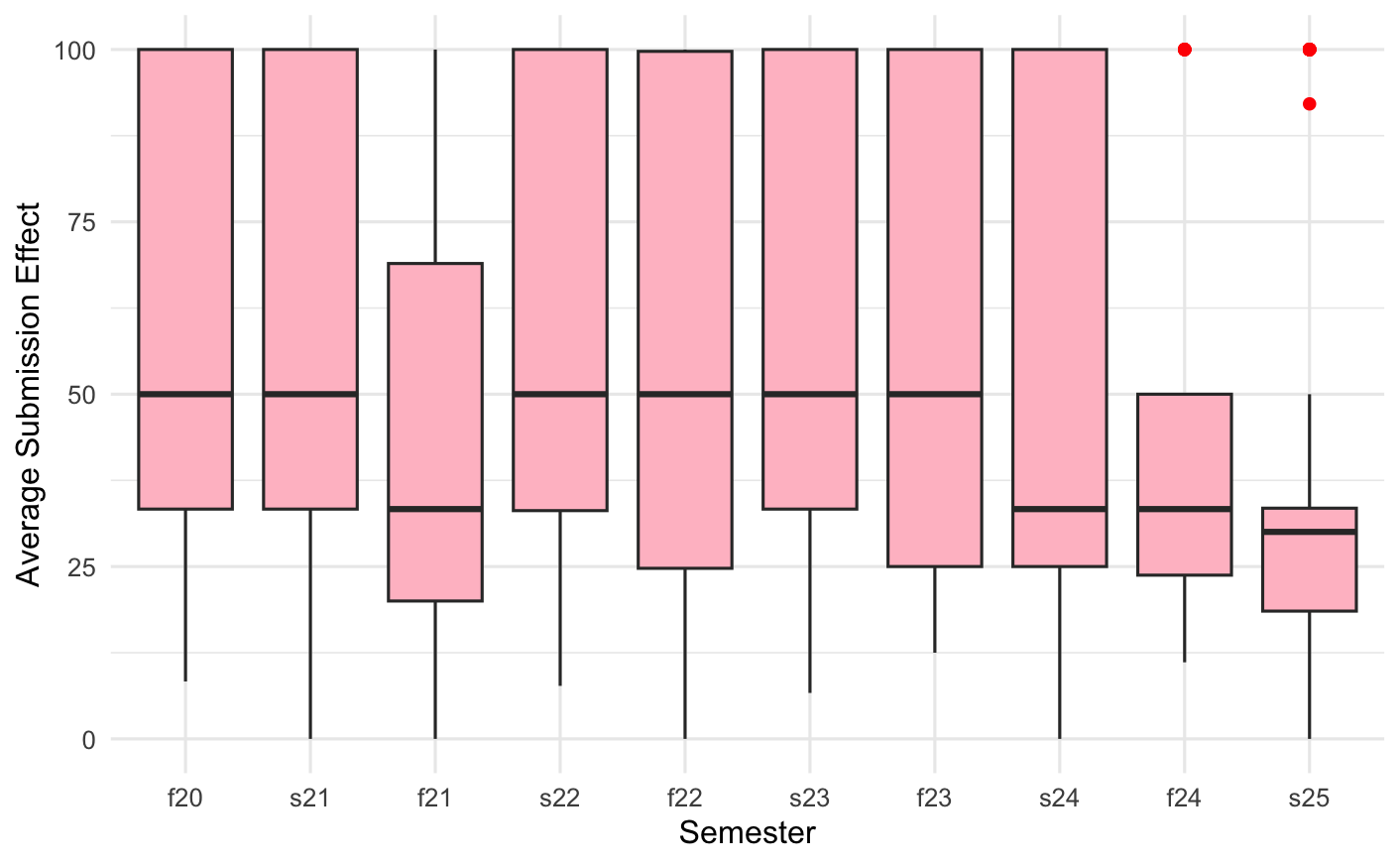}
        \dcaption{Average submission effect}
        \label{fig:avg_edit_effectiveness}
\end{figure}

\begin{figure}[tb]
  \centering
  \includegraphics[width=\linewidth]{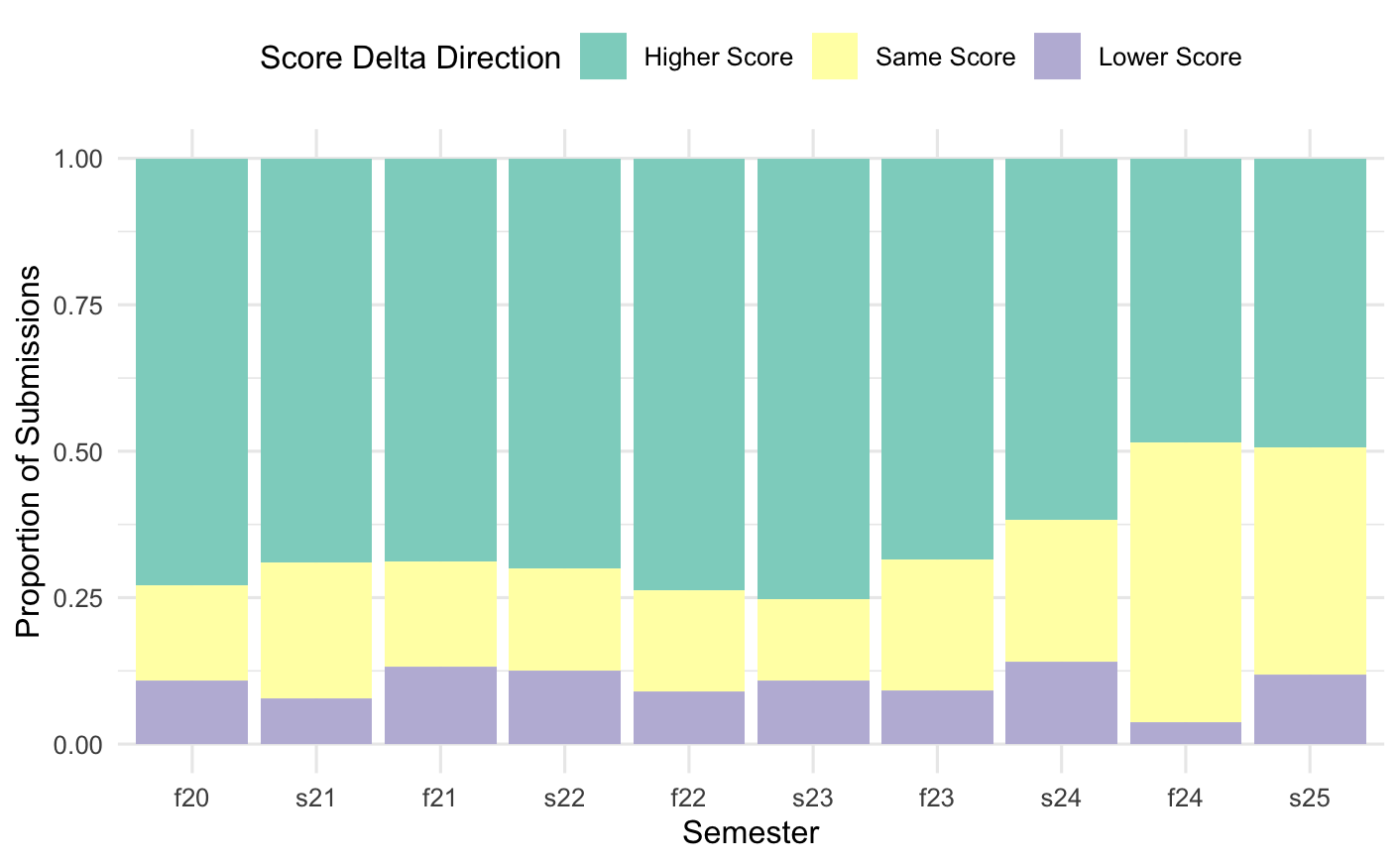}
  \dcaption{Submissions broken down by whether the score of the new submission increased, did not change, or decreased the score over the previous submission.}
  \label{fig:prop_submissions_score_deltas}        
\end{figure}

Since there is little variance in the final scores of the PageRank task to be explained by changing edit distances, we define another metric called the \texttt{Average Submission Effect}. It is the average change in the task score, measured by the sum of the changes between consecutive submissions, divided by the \texttt{Number of Submissions}. We observe a steady decrease in \texttt{Average Submission Effect} since s24, as shown in \fig{avg_edit_effectiveness}, suggesting that since s24 students make less improvements in their scores -- despite more edits in their code -- between consecutive submissions. The same change is observed in \fig{prop_submissions_score_deltas}, which shows that fewer submissions get a higher score than their preceding submission.

\begin{figure}[tb]
        \centering
        \includegraphics[width=\linewidth]{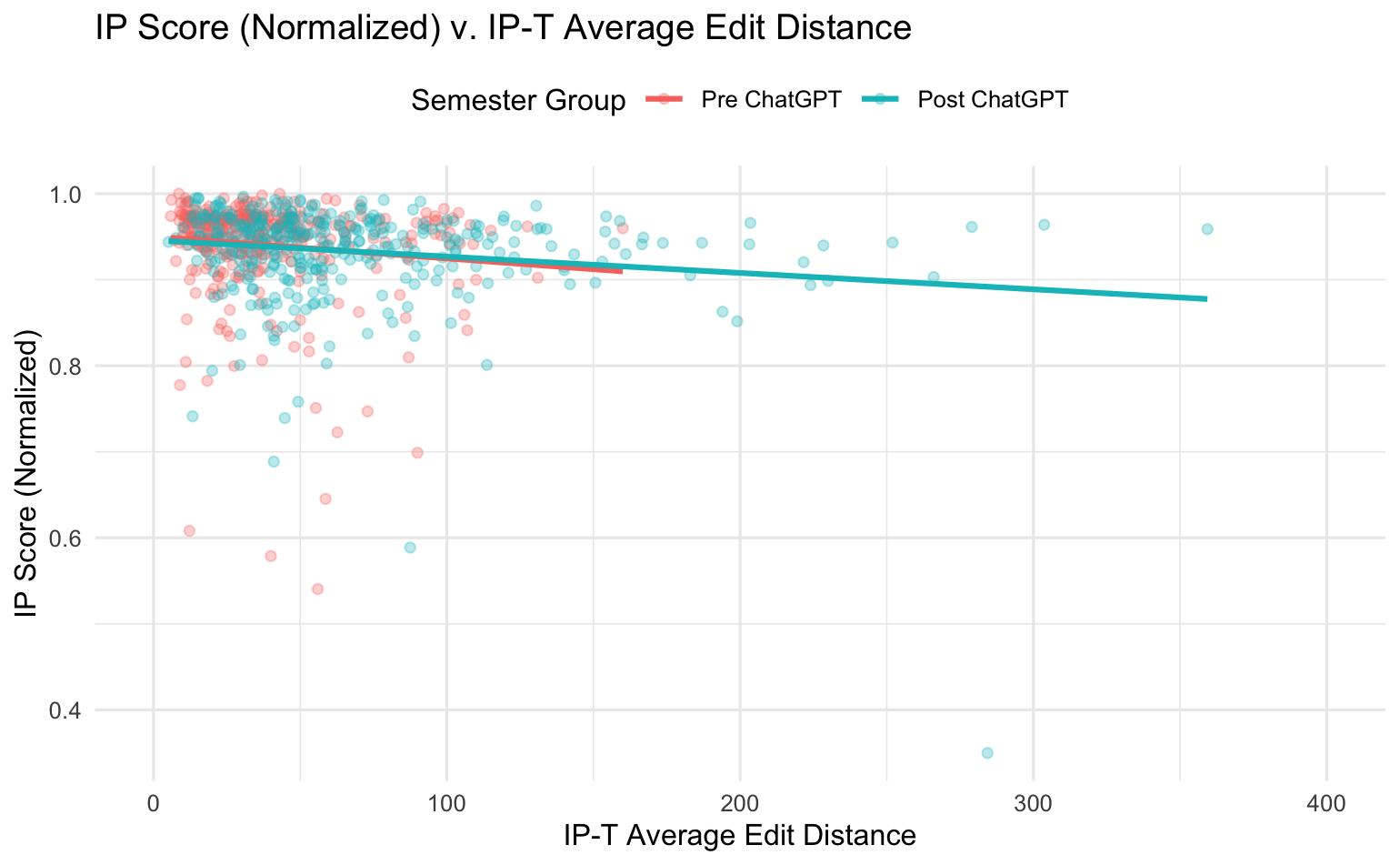}
        \dcaption{IP score v. PageRank task average edit distance}
        \label{fig:ip_score_vs_ipt_dist}
\end{figure}

There is also a negative correlation between \texttt{IP Score} and \texttt{Average Edit Distance} ($corr$ = -0.16, $p$ < 0.001), where for each additional 60 lines in \texttt{Average Edit Distance} for one's solution to the PageRank task, there is an associated
1\% decrease in one's \texttt{IP Score}, as shown in \fig{ip_score_vs_ipt_dist}. The
correlation also aligns with our hypothesis that more edits
between consecutive submissions suggests a less structured and incremental approach, and manifests in lower scores received for other individual projects as well. The correlation exist both Pre-ChatGPT and Post-ChatGPT.  

\begin{figure}[tb]
        \centering
        \includegraphics[width=\linewidth]{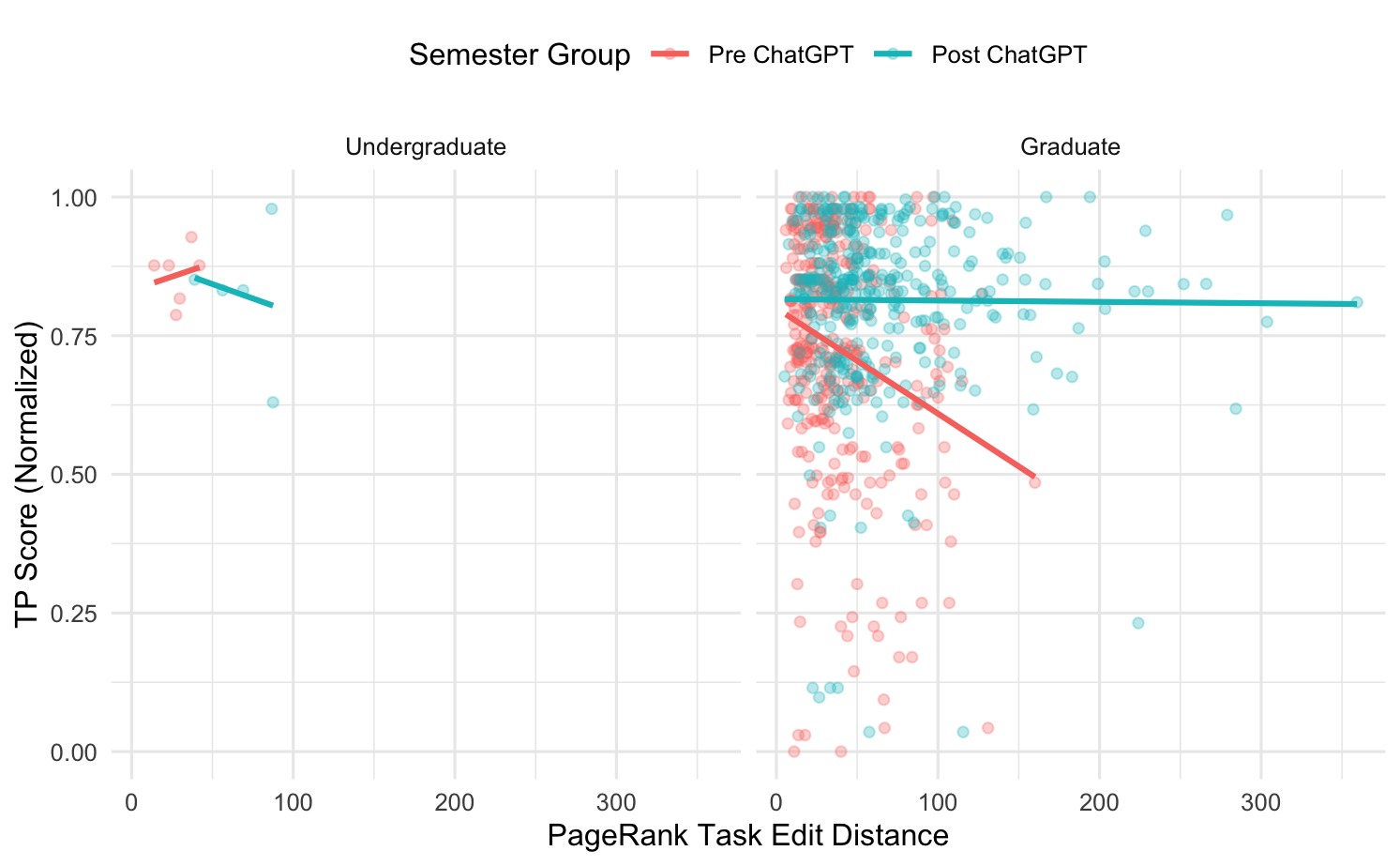}
        \dcaption{TP score v. PageRank task average edit distance}
        \label{fig:tp_v_avg_edit_dist}
\end{figure}


Pre-ChatGPT, higher edit distance in the PageRank task had a negative association with \texttt{TP Score}, but that association disappeared Post-ChatGPT, as shown in \fig{tp_v_avg_edit_dist}. One
possible explanation is that Pre-ChatGPT, both team project and
individual project scores have a similar relationship to Average Edit
Distances --- higher edits predict lower performance. But, the
availability of LLMs and the potential usage of them systematically
lifts the \texttt{Average Edit Distance}, flattening the
curve. Meanwhile, If a student does better in individual projects,
they are likely, though not guaranteed, to do better in the team
project. This is shown by a moderate, positive correlation ($corr$ =
0.240, $p$<0.001) between \texttt{IP Score}s and \texttt{TP Score}s.

\section{Discussions}\label{sect-discuss}

\textbf{Coding behavior changed significantly} since the introduction of LLMs, as observed in students' source code submissions for the PageRank task between f22 and s25, with
\begin{itemize}
    \item a 50\% increaes in median \texttt{Number of Submissions}
    \item a 300\% increase in median \texttt{Average Edit Distance}
    \item a 500\% increase in median \texttt{Total Edit Distance}
\end{itemize}

\noindent The trend aligns with our hypothesis that students have been submitting code that is generated by LLMs, even though it is against course policy. Our hypothesis is based on experiments that show that when LLMs are prompted to fix texts, they tend to introduce edits that are not prompted for, unless deliberately prompted against unnecessary changes and with a temperature setting of 0.

\textbf{Performance changed in mixed ways} over the same time period, with
\begin{itemize}
    \item no change in \texttt{Task Score}
    \item a small decrease in \texttt{IP Score} (Individual Projects Score)
    \item a moderate increase in \texttt{TP Score} (Team Project Score)
\end{itemize}

\noindent The combination of these observations may offer a pessimistic interpretation where
\begin{itemize}
    \item the task is simple enough to be solved by almost all students in all semesters, though students in post-ChatGPT semesters required more submissions and longer solutions
    \item the increasing edit distance is correlated with slightly lower individual projects (IP) scores, which suggests possibly lower learning outcomes.
    \item LLMs may be acting as additional members in teams, increasing their scores but masking individual contributions or disengagements.
\end{itemize}

\section{Limitations and Future Work}

\textbf{Speculating LLM utilization}: We do not know definitively to what extent, or if at all, any student utilized LLMs to generate code that they submitted for their programming projects. Although we have been surveying students on their LLM utilization since Spring 2023, students have an incentive to under-report as misuse may constitute an academic integrity violation.

\textbf{Sharing solutions among students}: Because the PageRank task has stayed the same over the years, students may have the means to acquire working solutions from other students who took the course in previous semesters. This study does not account for that. It may explain the decrease in edit distances prior to the introduction of ChatGPT.

\textbf{Generalizing to other programming projects}: While Scala (which is required for the PageRank task we analyzed) is widely used, there are many other programming languages with different syntax and conventions, and our findings may not generalize to them. Future work should examine whether the same observations exist in other programming projects within this course and within other courses.
\section{Conclusion}

The introduction of ChatGPT and other LLM-powered services has changed how students interact with programming assignments. Our multi-year study of source code submissions from a graduate-level cloud computing course reveals that while performance metrics such as individual project scores and team project scores have remained relatively stable, students’ coding behavior has changed in clearly observable ways since Fall 2022.

We observed a significant increase in edit distance (both on average and in total) and a greater number of submissions with negative score deltas, suggesting that students are modifying their submissions more extensively and perhaps less strategically. At the same time, submission effectiveness (measured by average score improvements between consecutive submissions) has declined. These findings are consistent with our belief that students are generating code with AI assistance. And these behaviors are correlated with minor but statistically significant drops in performance.

However, these correlations may be more nuanced. \texttt{IP Scores} decreased slightly, while \texttt{TP Scores} in fact increased. These opposing shifts may obscure deeper effects. Specifically, the growing gap between students’ individual capabilities and team performance points to an important side effect of LLM usage: the potential masking of individual learning gaps within team settings.

In the workforce context, the observed behavioral changes could potentially manifest as a growing tendency of junior developers to produce a larger number of commits that are also longer but less effective. This would likely result in increased requirements on companies' testing and quality assurance (QA) workflows in terms of both automated as well as human resources. Further, this may continually push the codebase towards increased size while at the same time decreasing its overall quality or maintainability, further exacerbating the costs involved in software development.

Our analysis shows that the emergence of LLM-powered development tools
has coincided with profound changes in how students approach programming tasks, and
these behavioral shifts are starting to show in student assessments. Students now make more frequent and larger edits, iterate
more often -- which we hypothesize comes from relying on AI systems for
implementation and debugging -- all without meaningful improvements in
their grades. This mismatch suggests that the current metrics we use
to evaluate learning are increasingly misaligned with the actual
skills needed in a world where intelligent tools are collaborators
rather than aids.

The centaur model may provide a powerful framework for reimagining
what and how we teach in this new landscape. In chess, the rise of
machine opponents did not end human play; it changed what counted as
skill. The most successful teams are those in which humans guide,
interpret, and synthesize machine output rather than compete with it
directly.  But, at the same time, the human team-member requires a
deep knowledge of chess.

We hope that the data presented here will provoke a thoughtful
discussion on how this model might apply to software engineering.  As
AI agents can assume more of the mechanical burden of coding, what is
the enduring value of the human developer?  What do students need to
learn to be effective members of a centaur-developer? How to teach them
what they need to learn?  Finally, how should we assess how students
are learning?  Just as chess mastery has been redefined in the age of
centaur play, so too must programming mastery be redefined---not as
the ability to produce code in isolation, but as the ability to build,
manage, and reason within powerful human-AI partnerships.

\begin{acks}

This research is funded in part by the Carnegie Mellon-Accenture Center of Excellence in AI-Enabled Workforce Training (ACE-AI). The Cloud Computing course at CMU is sponsored in part by Amazon Web Service, Google Cloud Platform, and Microsoft Azure.

\end{acks}

\bibliographystyle{ACM-Reference-Format}
\bibliography{0-lak26}

@Article{alanazi2025influence,
AUTHOR = {Alanazi, Manal and Soh, Ben and Samra, Halima and Li, Alice},
TITLE = {The Influence of Artificial Intelligence Tools on Learning Outcomes in Computer Programming: A Systematic Review and Meta-Analysis},
JOURNAL = {Computers},
VOLUME = {14},
YEAR = {2025},
NUMBER = {5},
ARTICLE-NUMBER = {185},
URL = {https://www.mdpi.com/2073-431X/14/5/185},
ISSN = {2073-431X},
DOI = {10.3390/computers14050185},
pages = {1}
}

@article{sun2024would,
  title={Would ChatGPT-facilitated programming mode impact college students’ programming behaviors, performances, and perceptions? An empirical study},
  author={Sun, Dan and Boudouaia, Azzeddine and Zhu, Chengcong and Li, Yan},
  journal={International Journal of Educational Technology in Higher Education},
  volume={21},
  number={1},
  pages={14},
  year={2024},
  publisher={Springer}
}

@article{jovst2024impact,
  title={The impact of large language models on programming education and student learning outcomes},
  author={Jo{\v{s}}t, Gregor and Taneski, Viktor and Karakati{\v{c}}, Sa{\v{s}}o},
  journal={Applied Sciences},
  volume={14},
  number={10},
  pages={4115},
  year={2024},
  publisher={MDPI}
}

@article{wang2025effect,
  title={The effect of ChatGPT on students’ learning performance, learning perception, and higher-order thinking: insights from a meta-analysis},
  author={Wang, Jin and Fan, Wenxiang},
  journal={Humanities and Social Sciences Communications},
  volume={12},
  number={1},
  pages={1--21},
  year={2025},
  publisher={Palgrave}
}

@inproceedings{lyu2024evaluating,
  title={Evaluating the effectiveness of llms in introductory computer science education: A semester-long field study},
  author={Lyu, Wenhan and Wang, Yimeng and Chung, Tingting and Sun, Yifan and Zhang, Yixuan},
  booktitle={Proceedings of the eleventh ACM conference on learning@ scale},
  publisher = {Association for Computing Machinery},
  address = {New York, NY, USA},
  pages={63--74},
  year={2024}
}

@inproceedings{ramirez2025understanding,
author = {Ramirez Osorio, Valeria and Zavaleta Bernuy, Angela and Simion, Bogdan and Liut, Michael},
title = {Understanding the Impact of Using Generative AI Tools in a Database Course},
year = {2025},
isbn = {9798400705311},
publisher = {Association for Computing Machinery},
address = {New York, NY, USA},
url = {https://doi.org/10.1145/3641554.3701785},
doi = {10.1145/3641554.3701785},
booktitle = {Proceedings of the 56th ACM Technical Symposium on Computer Science Education V. 1},
pages = {959–965},
numpages = {7},
location = {Pittsburgh, PA, USA},
series = {SIGCSETS 2025}
}

@article{kosar2024computer,
  title={Computer science education in ChatGPT era: Experiences from an experiment in a programming course for novice programmers},
  author={Kosar, Toma{\v{z}} and Ostoji{\'c}, Dragana and Liu, Yu David and Mernik, Marjan},
  journal={Mathematics},
  volume={12},
  number={5},
  pages={629},
  year={2024},
  publisher={MDPI}
}

@misc{prather2024widening,
      title={The Widening Gap: The Benefits and Harms of Generative AI for Novice Programmers}, 
      author={James Prather and Brent Reeves and Juho Leinonen and Stephen MacNeil and Arisoa S. Randrianasolo and Brett Becker and Bailey Kimmel and Jared Wright and Ben Briggs},
      year={2024},
      eprint={2405.17739},
      archivePrefix={arXiv},
      primaryClass={cs.AI},
      url={https://arxiv.org/abs/2405.17739}, 
}

@misc{becker2025measuring,
      title={Measuring the Impact of Early-2025 AI on Experienced Open-Source Developer Productivity}, 
      author={Joel Becker and Nate Rush and Elizabeth Barnes and David Rein},
      year={2025},
      eprint={2507.09089},
      archivePrefix={arXiv},
      primaryClass={cs.AI},
      url={https://arxiv.org/abs/2507.09089}, 
}

@inproceedings{Rasnayaka24,
author = {Rasnayaka, Sanka and Wang, Guanlin and Shariffdeen, Ridwan and Iyer, Ganesh Neelakanta},
title = {An Empirical Study on Usage and Perceptions of LLMs in a Software Engineering Project},
year = {2024},
isbn = {9798400705793},
publisher = {Association for Computing Machinery},
address = {New York, NY, USA},
url = {https://doi.org/10.1145/3643795.3648379},
doi = {10.1145/3643795.3648379},
booktitle = {Proceedings of the 1st International Workshop on Large Language Models for Code},
pages = {111–118},
numpages = {8},
location = {Lisbon, Portugal},
series = {LLM4Code '24}
}

@article{Khojah24,
author = {Khojah, Ranim and Mohamad, Mazen and Leitner, Philipp and de Oliveira Neto, Francisco Gomes},
title = {Beyond Code Generation: An Observational Study of ChatGPT Usage in Software Engineering Practice},
year = {2024},
issue_date = {July 2024},
publisher = {Association for Computing Machinery},
address = {New York, NY, USA},
volume = {1},
number = {FSE},
journal = {Proc. ACM Softw. Eng.},                  
url = {https://doi.org/10.1145/3660788},
doi = {10.1145/3660788},
month = jul,
articleno = {81},
numpages = {22}
}

@misc{chatgpt,
  author       = {OpenAI},
  title        = {Introducing ChatGPT},
  howpublished = {\url{https://openai.com/index/chatgpt/}},
  year         = {2022},
  note         = {Accessed: 2025-09-27}
}

@misc{myers,
  author       = {Git},
  title        = {git-diff Documentation: Diff Options},
  howpublished = {\url{https://git-scm.com/docs/diff-options/2.6.7}},
  year         = {2016},
  note         = {Accessed: 2025-09-27}
}

@inproceedings{xue24,
author = {Xue, Yuankai and Chen, Hanlin and Bai, Gina R. and Tairas, Robert and Huang, Yu},
title = {Does ChatGPT Help With Introductory Programming?An Experiment of Students Using ChatGPT in CS1},
year = {2024},
isbn = {9798400704987},
publisher = {Association for Computing Machinery},
address = {New York, NY, USA},
url = {https://doi.org/10.1145/3639474.3640076},
doi = {10.1145/3639474.3640076},
booktitle = {Proceedings of the 46th International Conference on Software Engineering: Software Engineering Education and Training},
pages = {331–341},
numpages = {11},
location = {Lisbon, Portugal},
series = {ICSE-SEET '24}
}

@inproceedings{Nguyen24,
author = {Nguyen, Sydney and Babe, Hannah McLean and Zi, Yangtian and Guha, Arjun and Anderson, Carolyn Jane and Feldman, Molly Q},
title = {How Beginning Programmers and Code LLMs (Mis)read Each Other},
year = {2024},
isbn = {9798400703300},
publisher = {Association for Computing Machinery},
address = {New York, NY, USA},
url = {https://doi.org/10.1145/3613904.3642706},
doi = {10.1145/3613904.3642706},
booktitle = {Proceedings of the 2024 CHI Conference on Human Factors in Computing Systems},
articleno = {651},
numpages = {26},
location = {Honolulu, HI, USA},
series = {CHI '24}
}

@misc{ray2025review,
title={A Review on Vibe Coding: Fundamentals, State-of-the-art, Challenges and Future Directions},
url={http://dx.doi.org/10.36227/techrxiv.174681482.27435614/v1},
DOI={10.36227/techrxiv.174681482.27435614/v1},
publisher={Institute of Electrical and Electronics Engineers (IEEE)},
author={Ray, Partha Pratim},
year={2025},
month=may }

@inproceedings{liffiton2023codehelp,
  title={Codehelp: Using large language models with guardrails for scalable support in programming classes},
  author={Liffiton, Mark and Sheese, Brad E and Savelka, Jaromir and Denny, Paul},
  booktitle={Proceedings of the 23rd Koli Calling International Conference on Computing Education Research},
  publisher = {Association for Computing Machinery},
  address = {New York, NY, USA},  pages={1--11},
  year={2023}
}

@inproceedings{sheese2024patterns,
  title={Patterns of student help-seeking when using a large language model-powered programming assistant},
  author={Sheese, Brad and Liffiton, Mark and Savelka, Jaromir and Denny, Paul},
  publisher = {Association for Computing Machinery},
  doi = {10.1145/3636243.3636249},
  booktitle={Proceedings of the 26th Australasian computing education conference},
  address = {New York, NY, USA},
  pages={49--57},
  year={2024}
}

@misc{sl25,
      title={Modeling the Centaur: Human-Machine Synergy in Sequential Decision Making}, 
      author={Shoresh, David and Loewenstein, Yonatan},
      year={2025},
      eprint={2412.18593},
      archivePrefix={arXiv},
      primaryClass={cs.HC},
      url={https://arxiv.org/abs/2412.18593}, 
}

@article{dk21,
  title     = {AI Should Augment Human Intelligence, Not Replace It},
  author    = {De Cremer, David and Kasparov, Garry},
  journal   = {Harvard Business Review},
  year      = {2021},
  month     = {March 18},
  volume={18},
  number={1},
  pages={1--8},
  url       = {https://hbr.org/2021/03/ai-should-augment-human-intelligence-not-replace-it}
}

\end{document}